\documentclass[letterpaper, 10 pt, conference]{ieeeconf}
\IEEEoverridecommandlockouts

\usepackage[utf8]{inputenc}
\usepackage{amsmath}
\usepackage{mathrsfs}
\usepackage{amssymb}
\usepackage{color}
\usepackage{hhline}
\usepackage{dsfont}
\usepackage{cite}
\usepackage{comment}
\usepackage{float}
\usepackage{graphicx}
\usepackage{BernsteinStyle}
\usepackage{float}
\usepackage{tikz}
\usepackage{booktabs}
\usepackage{tabularx}
\usepackage{multirow}
\usepackage[version=4]{mhchem}
\usepackage{boldline}
\usepackage{bm}
\usepackage{makecell}
\usepackage{diagbox}
\usepackage{color, colortbl}
\definecolor{Yellow}{rgb}{1,1,0}

\usepackage[colorinlistoftodos,textwidth=1 in,textsize=footnotesize]{todonotes}

\usetikzlibrary{shapes,arrows,calc,positioning}
\tikzstyle{bigblock} = [draw, fill=blue!20, rectangle, 
    minimum height=6em, minimum width=8em]
\tikzstyle{medblock} = [draw, fill=blue!20, rectangle, 
    minimum height=4em, minimum width=4em]    
\tikzstyle{mux} = [draw, fill=black!20, rectangle, 
    minimum height=5em, minimum width=0.1em]    
\tikzstyle{smallblock} = [draw, fill=blue!20, rectangle, 
    minimum height=3em, minimum width=4em]
\tikzstyle{sum} = [draw, fill=blue!20, circle, node distance=1cm]
\tikzstyle{signal} = [coordinate]
\tikzstyle{pinstyle} = [pin edge={to-,thin,black}]
\tikzstyle{block} = [draw, fill=blue!20, rectangle, 
    minimum height=3em, minimum width=6em]
\tikzstyle{blockS} = [draw, fill=blue!20, rectangle, 
    minimum height=3em, minimum width=4em]  
\tikzstyle{sum} = [draw, fill=blue!20, circle, node distance=1.5cm]
\tikzstyle{gain} = [draw, fill=blue!20, regular polygon, regular polygon sides = 3, node distance=1.25cm, shape border rotate = -90]
\tikzstyle{mult} = [draw, fill=blue!20, circle, node distance=1.25cm ,inner sep=0pt, minimum size = 0.3cm]

\tikzstyle{input} = [coordinate]
\tikzstyle{output} = [coordinate]

\newcounter{example}

\usepackage{hyperref}
\usepackage{xcolor}
\hypersetup{
    colorlinks,
    linkcolor={blue!100!black},
    citecolor={blue!50!black},
    urlcolor={blue!80!black}
}

\title{Adaptive Numerical Differentiation\\ for Extremum Seeking with Sensor Noise} 
\title{Extremum Seeking Control with Adaptive Numerical Differentiation to Mitigate Sensor Noise}

\title{Sensor-Noise Mitigation in Extremum Seeking Control\\ Using Adaptive Numerical Differentiation}

\author{\large Shashank Verma, Juan Augusto Paredes Salazar, Jhon Manuel Portella Delgado, \\ Ankit Goel and Dennis S. Bernstein
\thanks{Shashank Verma and Dennis S. Bernstein are with the Department of Aerospace Engineering, University of Michigan, Ann Arbor, MI, USA. {\tt\small \{shaaero, dsbaero\}@umich.edu}}
\thanks{Juan Augusto Paredes Salazar, Jhon Manuel Portella and Ankit Goel are with the Department of Mechanical Engineering, University of Maryland, Baltimore County, MD 21250.
{\tt \small \{japarede, jportel1, ankgoel\}@umbc.edu}}
}

\begin{document}

\maketitle

\begin{abstract}
Extremum-seeking control (ESC) is widely used to optimize performance when the system dynamics are uncertain.
However, sensitivity to sensor noise is a crucial issue in ESC implementation due to the use of high-pass filters or gradient estimators.
To reduce the sensitivity of ESC to noise, this paper investigates the use of the recently developed adaptive input and state estimation (AISE) technique for numerical differentiation.
In particular, this paper develops extremum-seeking control with adaptive input and state estimation (ESC/AISE), where AISE replaces the high-pass filter of ESC to improve performance under sensor noise.
The effectiveness of ESC/AISE is illustrated via numerical examples.
\end{abstract}

\section{Introduction}\label{sec:introduction}


Extremum-seeking control (ESC) is a real-time, model-free optimization method for tuning an input to maximize or minimize a performance variable in the presence of large modeling uncertainty and unmodeled dynamics. 
The central idea is to inject a small periodic perturbation, called a dither signal, into the input, measure the resulting performance response, and use demodulation and filtering to form an estimate of the local gradient.
Under standard assumptions, including the smoothness of the steady-state input–output map, stability of the plant dynamics, and sufficient separation between the dither frequency and the plant bandwidth, averaging theory demonstrates that the closed loop behaves like a gradient method and converges to a neighborhood of a local optimizer \cite{KrsticBookESC2003,scheinker2024}.
The approach has been validated across diverse domains, including robotics  \cite{matveev2015,bagheri2018}, energy management \cite{ghaffari2014,zhou2018}, combustion  \cite{banaszuk2004,liu2023}, and nuclear fusion \cite{lanctot2016,dubbioso2024}.


Despite extensive empirical success, ESC can exhibit performance limitations, such as narrow stability margins, sensitivity to operating-point drift, and degraded transients, in certain settings \cite{krstic2000}. 
To enhance robustness and dynamic performance, numerous variants have been proposed, including refined demodulation/filtering schemes, adaptive dither scheduling, multi-parameter and higher-order updates, and learning- or robustness-augmented formulations \cite{liu2010,ghaffari2012,gelbert2012,scheinker2016,mele2021,guay2021,williams2024,juanQSRCAC2024,juanRCESC2024}. 
A persistent challenge is measurement noise;  since gradient estimates are formed via demodulation and derivative-/high-pass-like operations, sensor noise is readily amplified.
Accordingly, several modifications have been implemented to mitigate the sensitivity to sensor noise \cite{stankovic2010_2,brinon2013,atanasov2015,wu2015,liu2016,radenkovic2018,scheinker2021,sadatieh2021,zhao2023,yang2024,dewasme2024}. 
However, noise sensitivity remains a key practical issue in ESC implementations, especially when high-pass filters or gradient estimators are used.

In this paper, we propose a discrete-time extremum-seeking controller that improves optimization performance under sensor noise by replacing the conventional high-pass filter with \textit{adaptive input and state estimation} (AISE). The resulting scheme, called ESC/AISE, uses AISE as a real-time, causal, model-independent numerical differentiator that adapts to changing noise statistics, thereby mitigating the noise amplification commonly associated with demodulation and high-pass/derivative operations \cite{verma_shashank_ACC2022,verma_shashank_2024_realtime_IJC,verma_shashank_2024_realtime_VRF_ACC,verma_shashank_verma2025realtime_VRF_CEP}. 
In this paper, we focus on the SISO case for clarity. 
Further details on AISE and comparisons with alternative real-time numerical differentiation methods are described in \cite{verma_shashank_2024_realtime_IJC}.




The contents of the paper are as follows.
Section \ref{sec:control_statement} provides a statement of the control problem, which involves continuous-time dynamics under sampled-data feedback control.
Section \ref{sec:ESC} provides a review of discrete-time ESC.
Section \ref{sec:AISE_ESC} introduces ESC/AISE, where the high-pass filter in ESC is replaced by AISE.
Section \ref{sec:numerical_examples} presents examples that illustrate the performance of ESC/AISE when sensor noise is added to the system output and compares it against discrete-time ESC.
Finally, Section \ref{sec:conclusions} presents conclusions.

{\bf Notation:}
$\BBR \isdef (-\infty,$ $\infty),$ 
$\BBC$ denotes the complex numbers,
$\Vert\cdot\Vert$ denotes the Euclidean norm on $\BBC^n,$ and 
$\bfz\in\BBC$ denotes the Z-transform variable.
$I_n$ denotes an $n \times n$ identity matrix.
$\lfloor \cdot \rfloor$ denotes the floor function.
For all $x\in\BBR$ and  $\varepsilon > 0,$ $\BBB_\varepsilon (x)$ denotes the open ball of radius $\varepsilon$ centered at $x$.
%


\section{Problem Statement}\label{sec:control_statement}

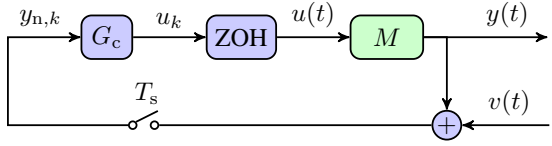
\begin{figure} [h!]
    \centering
    \resizebox{0.85\columnwidth}{!}{%
    \begin{tikzpicture}[>={stealth'}, line width = 0.25mm]

    \node [input, name=ref]{};
    \node [smallblock, rounded corners, right = 0.5cm of ref , minimum height = 0.6cm, minimum width = 0.7cm] (controller) {$G_\rmc$};
    \node [smallblock, rounded corners, right = 1cm of controller, minimum height = 0.6cm , minimum width = 0.5cm] (DA) {ZOH};
    
    \node [smallblock, fill=green!20, rounded corners, right = 1cm of DA, minimum height = 0.6cm , minimum width = 1cm] (system) {$M$};
    \node [output, right = 1.7cm of system] (output) {};
    \node [input, below = 0.9cm of system] (midpoint) {};

    \node[sum, inner sep = 0.4em, right = 1.75em of midpoint.east] (sum_d) {};
    \node[draw = none] at (sum_d.center) {$+$};
    
    \draw [->] (controller) -- node [above] {$u_k$} (DA);
    \draw [->] (DA) -- node [above] {$u (t)$} (system);
    \draw [<-] (sum_d.east) -- node [above, xshift = 0.05cm] {$v (t)$} ([xshift = 1.2cm]sum_d.east);
    
    \node[circle,draw=black, fill=white, inner sep=0pt,minimum size=3pt] (rc11) at ([xshift=-3.2cm]midpoint) {};
    \node[circle,draw=black, fill=white, inner sep=0pt,minimum size=3pt] (rc21) at ([xshift=-3.5cm]midpoint) {};
    \draw [-] (rc21.north east) --node[below,yshift=.6cm]{$T_\rms$} ([xshift=.3cm,yshift=.15cm]rc21.north east) {};
    
    \draw [->] (system) -- node [name=y, near end]{} node [above, xshift = 0.3cm] {$y (t)$}(output);
    
    \draw [->] (system.east) -| (sum_d.north);
    \draw [-] (sum_d.west) -| (rc11.east);
    \draw [->] (rc21) -| ([xshift = -1cm]controller.west) -- node [near end, above, xshift = -0.3cm] {$y_{\rmn, k}$} (controller.west);
    
    \end{tikzpicture}
    }  
    \caption{Sampled-data implementation of the discrete-time controller $G_\rmc$ for controlling the continuous-time system $M$ with input $u,$ output $y,$ and sensor noise $v.$
    All sample-and-hold operations are synchronous, and the sampling time is given by $T_\rms>0.$
    The discrete-time controller uses the sampled noisy measurement $y_{\rmn, k} \isdef y(k T_\rms) + v(k T_\rms)$ as the input and generates the discrete-time control $u_k$ at each step $k$.
    The resulting continuous-time control $u(t)$ is generated by applying a zero-order-hold operation to $u_k$.
    The objective of the controller is to provide an input $u(t)$ that converges to a neighborhood of an input value that either locally minimizes or locally maximizes the output $y(t).$
    }
    \label{fig:AC_CT_blk_diag}
\end{figure}

We consider continuous-time dynamics under sampled-data control using discrete-time control to reflect the practical implementation of digital controllers for physical systems.
In particular, we consider the control architecture in Figure \ref{fig:AC_CT_blk_diag}, where $M$ is the target continuous-time system, for all $t\ge 0$, $u(t)\in\BBR$ is the control, $y(t)\in\BBR$ is the output of $M,$ and $v(t)\in\BBR$ is the sensor noise.

The noisy measurements $y_{\rmn, k} $ are generated by sampling the output $y(t)$ corrupted by the noise signal $v(t),$ that is
\begin{align}
    y_{\rmn, k} &\isdef y(k T_\rms) + v (k T_\rms) =  y_k + v_k,
\end{align}
where $T_\rms>0$ is the sampling time.
The discrete-time controller is denoted by $G_\rmc$.
The input to $G_\rmc$ is $y_{\rmn, k}$, and its output at each step $k$ is the discrete-time control $u_k\in\BBR.$
The continuous-time control $u(t)$ applied to the structure is generated by applying a zero-order-hold operation to $u_k,$ that is,
for all $k\ge0,$ and, for all $t\in[kT_\rms, (k+1) T_\rms),$ 
\begin{equation}
    u(t) = u_k.
\end{equation}









Let $\mathcal{U}_{\rm min} \subseteq \BBR$ be the set of values of $u$ that locally minimize $y,$ and let $\mathcal{U}_{\rm max} \subseteq \BBR$ be the set of values of $u$ that locally maximize $y.$
Note that $u$ locally minimizes $y$ if and only if $u$ locally maximizes $-y$.
We assume that $\mathcal{U}_{\rm min}$ and $\mathcal{U}_{\rm max}$ have no accumulation points.
The objective of the discrete-time controller is to provide an input $u(t)$ such that the output  $y(t)$ converges to a neighborhood of either a local minimizer or a local maximizer.
When the objective is minimization, the objective is to obtain $u(t)$ such that there exist $\varepsilon > 0,$ $t_\rmc > 0,$ and $u_{\rm min}^\star \in \mathcal{U}_{\rm min}$ such that $\BBB_\varepsilon (u_{\rm min}^\star) \cap \mathcal{U}_{\rm min} = \{u_{\rm min}^\star\}$ and, for all $t \ge t_\rmc,$ $u(t) \in \BBB_\varepsilon (u_{\rm min}^\star).$
%
%
%
When the objective is maximization, the objective is to obtain $u(t)$ such that there exist $\varepsilon > 0,$ $t_\rmc > 0,$ and $u_{\rm max}^\star \in \mathcal{U}_{\rm max}$ such that $\BBB_\varepsilon (u_{\rm max}^\star) \cap \mathcal{U}_{\rm max} = \{u_{\rm max}^\star\}$ and, for all $t \ge t_\rmc,$ $u(t) \in \BBB_\varepsilon (u_{\rm max}^\star).$
%

\section{Overview of Discrete-Time Extremum-seeking control}\label{sec:ESC}

This section provides an overview of the discrete-time ESC scheme shown in\cite[ch.~4.1]{KrsticBookESC2003} with an added low-pass filter, as suggested in \cite[ch.~5.2]{KrsticBookESC2003}, and an enabling gain.
For all $k\ge1,$ the update equations for discrete-time ESC are given by
\begin{align}
y_{\rmh, k} &= -\omega_\rmh T_\rms y_{\rmh, k-1} + y_{\rmg, k} - y_{\rmg, k-1}, \label{eq:yhk1}\\
y_{\rml, k} &= (1 - \omega_\rml T_\rms) y_{\rml, k-1} \nn \\
& \quad + \omega_\rml T_\rms K_{{\rm en}, k} y_{\rmh, k-1} A_{\rm esc} \sin{(\omega_{\rm esc} T_\rms (k-1))}, \label{eq:ylk}\\
y_{{\rm esc}, k} &= y_{{\rm esc}, k-1} + y_{\rml, k-1}, \label{eq:yesck}\\
u_k &= K_{\rm esc} y_{{\rm esc}, k} + A_{\rm esc} \sin{(\omega_{\rm esc} T_\rms k)} + u_0, \label{eq:uesc}
\end{align}
where $y_{\rmg, k} \isdef K_\rmg y_{\rmn, k},$ $K_\rmg > 0$ is a scaling gain, $y_{\rmh, k},$ $y_{\rml, k},$ $y_{{\rm esc}, k}\in\BBR$ are internal states, $K_{{\rm en}, k} \in \{0, 1\}$ is an enabling gain, $K_{\rm esc}$ is the ESC output gain, $\omega_\rml > 0$ is the cutoff frequency of the low-pass filter, $\omega_\rmh > 0$ is the cutoff frequency of the high-pass filter, $u_0\in\BBR$ is the control input bias term, $A_{\rm esc}, \omega_{\rm esc} > 0$ are the amplitude and frequency of the ESC perturbation signal, respectively, and $u_k\in\BBR$ is the ESC output.

When $K_{{\rm en}, k} = 0,$ $y_{\rmh, k}$ is updated, and $y_{\rml, k}$ converges to 0.
In  the case where $K_{{\rm en}, k} = 1,$ $y_{\rmh, k},$ $y_{\rml, k},$ and $y_{{\rm esc}, k}$ are all updated.
Consequently, $K_{{\rm en}, k} = 0$ can be used to stop $y_{{\rm esc}, k}$ from updating when an adequate optimizer is reached.
Furthermore, when the objective is minimization, $K_{\rm esc} < 0,$ whereas, when the objective is maximization, $K_{\rm esc} > 0.$
The block diagram for discrete-time ESC is shown in Figure \ref{fig:DT_ESC}.

\begin{figure*}[!ht]
\vspace{1em}
\centering
\begin{tikzpicture}[>={stealth'}, line width = 0.25mm]
\node[draw = none] at (0,0) (orig) {};
\node [smallblock, rounded corners, minimum height = 3.25cm, minimum width = 12.9cm] at ([yshift = -0.25em, xshift = -2.25em]orig.center) (esc_controller) {};
\node[below right] at (esc_controller.north west) {$G_\rmc$ (ESC)};
\node[smallblock, fill=red!20, rounded corners, minimum height = 3em, minimum width = 3em] at (orig.center) (LP) {\Large$\frac{\omega_\rml T_\rms}{\bfz + \omega_\rml T_\rms - 1}$};
\node[smallblock, fill=red!20, rounded corners, minimum height = 2em, minimum width = 2em, left = 1em of LP.west] (Ken) {$K_{{\rm en}, k}$};
\node[sum, fill=red!20, inner sep = 0.4em, left = 1em of Ken.west] (mult_c) {};
\node[draw = none] at (mult_c.center) {$\times$};
\node[smallblock, fill=green!20, rounded corners, minimum height = 3em, minimum width = 3em, left = 2.25em of mult_c.west] (HP) {\Large$\frac{\bfz - 1}{\bfz + \omega_\rmh T_\rms}$};
\node[smallblock, fill=red!20, rounded corners, minimum height = 2em, minimum width = 2em, left = 2.25em of HP.west] (Kgain) {$K_{\rmg}$};
\node[smallblock, fill=red!20, rounded corners, minimum height = 3em, minimum width = 3em, right = 2.25em of LP.east] (ESC_int) {\Large$\frac{1}{\bfz - 1}$};
\node[smallblock, fill=red!20, rounded corners, minimum height = 2em, minimum width = 2em, right = 2.75em of ESC_int.east] (Kesc) {$K_{\rm esc}$};
\node[sum, fill=red!20, inner sep = 0.4em, right = 1em of Kesc.east] (sum_c) {};
\node[draw = none] at (sum_c.center) {$+$};
\node[draw = none] at ([xshift = -6em, yshift = -3.5em]mult_c.center) (sineESC) {$A_{\rm esc} \sin{(\omega_{\rm esc} T_\rms k})$};
\node[draw = none] at ([yshift = 2.5em]sum_c.center) (omega0) {$u_0$};
\draw[->] (sineESC.east) -| (sum_c.south);
\draw[->] (sineESC.east) -| (mult_c.south);
\draw[->] (omega0.south) -- (sum_c.north);
\draw[->] (Kgain.east) -- node[above] {$y_{\rmg, k}$} (HP.west);
\draw[->] (HP.east) -- node[above] {$y_{\rmh, k}$} (mult_c.west);
\draw[->] (mult_c.east) -- (Ken.west);
\draw[->] (Ken.east) -- (LP.west);
\draw[->] (LP.east) -- node[above] {$y_{\rml, k}$} (ESC_int.west);
\draw[->] (ESC_int.east) -- node[above] {$y_{{\rm esc}, k}$} (Kesc.west);
\draw[->] (Kesc.east) -- (sum_c.west);
\draw[->] ([xshift = -3em]Kgain.west) -- node[above, xshift = -0.5em] {$y_{\rmn, k}$}(Kgain.west);
\draw[->] (sum_c.east) -- node[above, xshift = 0.4em] {$u_k$}([xshift = 2em]sum_c.east);
\end{tikzpicture}
\caption{ 
Discrete-time extremum-seeking control described by \eqref{eq:yhk1}-\eqref{eq:uesc}.
Note that $\bfz$ is the forward-shift operator.
}
\label{fig:DT_ESC}
\end{figure*}

\section{Extremum-Seeking Control with Adaptive Input and State Estimation}\label{sec:AISE_ESC}

This section formulates the ESC/AISE algorithm.
In particular, Subsection \ref{subsec:AISE} provides a brief overview AISE for numerical differentiation,
and Subsection \ref{subsec:AISE_ESC} introduces ESC/AISE, where the ESC high-pass filter \eqref{eq:yhk1} described in Section \ref{sec:ESC} is replaced by AISE.

\subsection{Overview pf Adaptive Input and State Estimation for Numerical Differentiation}\label{subsec:AISE}

This section provides a brief overview of the adaptive input and state estimation (AISE) algorithm
The theory of the adaptive input and state estimation (AISE) algorithm is described in detail in Sections $4$ and $5$ in \cite{verma2024real}, and in Section $4$ in \cite{verma2025real}.
The theory of the AISE algorithm is described in detail in Sections $4$ and $5$ in \cite{verma2024real}, and in Section $4$ in \cite{verma2025real}.
%
%
Furthermore, the application of AISE for real-time numerical differentiation for SISO systems is described in detail in \cite{verma_shashank_ACC2022,verma_shashank_2024_realtime_IJC, verma_shashank_2024_realtime_VRF_ACC, verma_shashank_verma2025realtime_VRF_CEP}.

Consider the linear discrete-time SISO system
%
\begin{align}
	x_{k+1} &=  A x_{k} + Bd_{k}, 	\label{state_eqn}\\
	y_k  &= C x_k + {D} v_k, \label{output_eqn}
\end{align}
where
$k\ge0$ is the step,
$x_k \in \mathbb R$ is the unknown state,
$d_k \in \mathbb R$ is an unknown input,
$y_k \in \mathbb R$ is a measured output,
$v_k \in \mathbb R$ is standard white noise, 
${D}v_k \in \mathbb R$ is the sensor noise at time $t = kT_\rms$, where $T_\rms$ is the sample time, and ${D}$ is assumed to be unknown. 

In the AISE framework for real-time numerical differentiation, \eqref{state_eqn} and \eqref{output_eqn} model a discrete-time integrator.
For example, for single discrete-time differentiation, $A = 1$, $B = T_\rms$, and $C = 1,$ that is, $y_k$ is obtained by integrating $d_k.$
Conversely, $d_k$ is the first derivative of $y_k.$
In the noise-free case, that is ${D}=0,$ note that $y_k = x_k,$ and $d_k = \frac{x_{k+1} - x_k}{T_s},$ which is the derivative of the sampled output $y_k.$
%

%
The AISE algorithm estimates the unknown state $x_k$ and the unknown input $d_k.$
The details of the algorithm can be found in \cite{verma2024real}.
In the context of numerical differentiation, with the appropriate choice of $A,B, C, D$, AISE provides an estimate $\hat{d}_k$ of the unknown input $d_k,$ which is the derivative of the sampled output $y_k.$
The parameters used in the AISE algorithm, which are specified in the Examples in Section \ref{sec:numerical_examples}, are summarized in Table \ref{tab:AISE_param_table}.
Brief guidelines for parameter tuning are provided next:
\begin{itemize}
    \item $\bm{n_\rme :}$ Increasing this parameter increases the amount of data considered in the estimation of $d_k$ and can yield more accurate values of $\hat{d}_k.$
    \item $\bm{n_\rmf :}$ Increasing this parameter increases the order of the target model in the $d_k$ estimation procedure and allows $\hat{d}_k$ to be more consistent with the model shown in \eqref{state_eqn} and \eqref{output_eqn}.
    \item $\bm{R_\theta :}$ Decreasing the norm of this parameter results in more aggressive adaptation during the initial adaptation, and can increase the speed of estimation of $d_k$, although this may destabilize the adaptation procedure.
    \item $\bm{R_z, R_d :}$ Decreasing the norms of these parameters results in a more aggressive adaptation in response to large norm values of the estimation error and $\hat{d}_k,$ respectively.
    \item $\bm{R_\infty :}$ This parameter determines an upper value for the covariance in the adaptation update equation, which prevents the covariance from growing unbounded; \cite{lai2022exponential} provides a more detailed explanation.
    \item $\bm{\tau_\rmn, \tau_\rmd, \eta, \alpha :}$ These parameters determine the extent to which the estimation algorithm adapts to new data samples by "forgetting" old data samples; \cite{mohseni2022recursive_2} provides a more detailed explanation
    \item $\bm{\eta_L, \eta_U, \beta}$: The range given by $[\eta_L, \eta_U]$ can be increased to improve the accuracy of the estimates of the measurement and process noise covariance matrices. $\beta$ is usually set to 0.5.
\end{itemize}

\begin{table}[ht]
    \centering
    \rowcolors{2}{gray!15}{white}
    \begin{tabularx}{\columnwidth}{|p{0.20\columnwidth}|X|}
    \hline
    \rowcolor{cyan!70!black}
    \textcolor{white}{Parameter} 
    & 
    \textcolor{white}{Definition}
    \\
    \hline
        $
            n_\rme
        $ 
        & 
        Input-estimation subsystem order.
    \\
    \hline
        $
            n_\rmf
        $
        &
        Order of the finite impulse response (FIR) filter $G_{f,k}$ in Section $4$ in \cite{verma2024real}.
    \\
    \hline
        $
            R_{\theta}
        $
        &
        Positive definite regularization weighting matrix.
    \\
    \hline
        $
            R_z,R_d
        $
        &
        Positive definite weighing matrices.
    \\
    \hline
        $
            R_\infty
        $
        &
        Resetting matrix used to prevent instability in the covariance matrix propagation procedure.
    \\
    \hline
        $
            \tau_\rmn, \tau_\rmd
        $
        &
        Short-term and Long-term sample sizes for the sample variance of the residual errors over the previous $\tau$ steps, as shown in Section $4$ in \cite{verma2025real}, such that $\tau_{d} >\tau_n > 0$
    \\
    \hline
        $
          \eta  \geq 0
        $
        &
        Tuning parameter for the selection of the forgetting factor, as described in Section $4$ in \cite{verma2025real}.
    \\
    \hline
        $
            \alpha  \in [0,1]
        $
        &
        Significance level associated with the cumulative F-distribution $F_{2\tau_n,b}^{-1},$ as shown in Section $4$ in \cite{verma2025real}.
    \\
    \hline
        $
            \eta_L,\eta_U
        $
        &
        Lower and Upper bounds for the decision variable $\eta$ such that $\eta_U\geq\eta_L\geq0.$
    \\
    \hline
        $
            \beta \in [0,1]
        $
        &
            Tuning parameter used 
            %
            %
            to define the optimization problem (61) in Proposition $4.1$ in \cite{verma2025real}. In most cases, its value is set to 0.5.
    \\
    \hline
    \end{tabularx}
    \caption{Parameters used in the implementation of AISE algorithm.}
    \label{tab:AISE_param_table}
\end{table}
\subsection{Extremum-Seeking Control with Adaptive Input and State Estimation}\label{subsec:AISE_ESC}

\begin{figure*}[h]
\vspace{1em}
\centering
\begin{tikzpicture}[>={stealth'}, line width = 0.25mm]
\node[draw = none] at (0,0) (orig) {};
\node [smallblock, rounded corners, minimum height = 3.25cm, minimum width = 12.8cm] at ([yshift = -0.25em, xshift = -1.8em]orig.center) (esc_controller) {};
\node[below right] at (esc_controller.north west) {$G_\rmc$ (ESC/AISE)};
\node[smallblock, fill=red!20, rounded corners, minimum height = 3em, minimum width = 3em] at (orig.center) (LP) {\Large$\frac{\omega_\rml T_\rms}{\bfz + (\omega_\rml T_\rms - 1)}$};
\node[smallblock, fill=red!20, rounded corners, minimum height = 2em, minimum width = 2em, left = 1em of LP.west] (Ken) {$K_{{\rm en}, k}$};
\node[sum, fill=red!20, inner sep = 0.4em, left = 1em of Ken.west] (mult_c) {};
\node[draw = none] at (mult_c.center) {$\times$};
\node[smallblock, fill=green!20, rounded corners, minimum height = 3em, minimum width = 3em, left = 2.25em of mult_c.west] (HP) { AISE };
\node[smallblock, fill=red!20, rounded corners, minimum height = 2em, minimum width = 2em, left = 2.25em of HP.west] (Kgain) {$K_{\rmg}$};
\node[smallblock, fill=red!20, rounded corners, minimum height = 3em, minimum width = 3em, right = 2.25em of LP.east] (ESC_int) {\Large$\frac{1}{\bfz - 1}$};
\node[smallblock, fill=red!20, rounded corners, minimum height = 2em, minimum width = 2em, right = 2.75em of ESC_int.east] (Kesc) {$K_{\rm esc}$};
\node[sum, fill=red!20, inner sep = 0.4em, right = 1em of Kesc.east] (sum_c) {};
\node[draw = none] at (sum_c.center) {$+$};
\node[draw = none] at ([xshift = -6em, yshift = -3.5em]mult_c.center) (sineESC) {$A_{\rm esc} \sin{(\omega_{\rm esc} T_\rms k})$};
\node[draw = none] at ([yshift = 2.5em]sum_c.center) (omega0) {$u_0$};
\draw[->] (sineESC.east) -| (sum_c.south);
\draw[->] (sineESC.east) -| (mult_c.south);
\draw[->] (omega0.south) -- (sum_c.north);
\draw[->] (Kgain.east) -- node[above] {$y_{\rmg, k}$} (HP.west);
\draw[->] (HP.east) -- node[above] {$y_{\rmh, k}$} (mult_c.west);
\draw[->] (mult_c.east) -- (Ken.west);
\draw[->] (Ken.east) -- (LP.west);
\draw[->] (LP.east) -- node[above] {$y_{\rml, k}$} (ESC_int.west);
\draw[->] (ESC_int.east) -- node[above] {$y_{{\rm esc}, k}$} (Kesc.west);
\draw[->] (Kesc.east) -- (sum_c.west);
\draw[->] ([xshift = -3.5em]Kgain.west) -- node[above, xshift = -0.5em] {$y_{\rmn, k}$}(Kgain.west);
\draw[->] (sum_c.east) -- node[above, xshift = 0.4em] {$u_k$}([xshift = 2em]sum_c.east);
\end{tikzpicture}
\caption{\footnotesize 
Discrete-time extremum-seeking control with adaptive input and state estimation (ESC/AISE).
}
\label{fig:AISE_ESC}
\end{figure*}
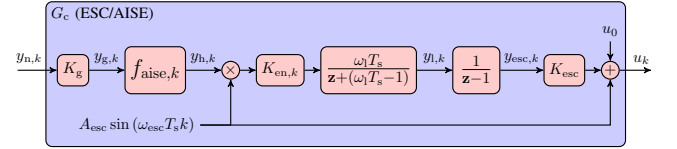

For all $k\ge1,$ let $y_{\rmh, k}$ be updated by
%
\begin{align}
     y_{\rmh, k} 
        &=
            \hat{d}_k(y_{\rmg, k}),
    \label{eq:yhk2}
\end{align}
where $\hat{d}_k$ is computed at each step $k,$ following the procedure presented in Section $4$ in \cite{verma2025real}.
%
%
Then, for all $k\ge1,$ the update equations for ESC/AISE are given by \eqref{eq:ylk}, \eqref{eq:yesck}, \eqref{eq:uesc}, \eqref{eq:yhk2}.
Note that \eqref{eq:yhk2} replaces \eqref{eq:yhk1}.
The block diagram for ESC/AISE is shown in Figure \ref{fig:AISE_ESC}.
%


\section{Numerical Examples}\label{sec:numerical_examples}

In this section, we investigate the performance of ESC/AISE and compare it with that of a similarly tuned discrete-time ESC in terms of performance and accuracy. 
%
To assess the accuracy of ESC/AISE, we define the root-mean-square error (RMSE) as
\begin{align}
{\rm RMSE} \isdef
\sqrt{ \frac{1}{k_{\rm end} - k_{\rm init}}\sum_{k=k_{\rm init}}^{k_{\rm end}}(u_k-u_{\rm opt})^2}, \label{rms}
\end{align}
where $[k_{\rm init}, k_{\rm end}]$ is the interval within which  RMSE is computed, and $u_{\rm opt}$ represents the optimal input that either minimizes or maximizes the measured output.
Note that $k_{\rm init}$ is chosen to exclude the transient response and thus consider only the steady-state response in the computation of the RMSE metric.

In Example \ref{ex:static_quadratic}, the objective is to minimize a quadratic function with sensor noise, which extends Example 1 of \cite{stankovic2010_2} to include sensor noise.
In Example \ref{ex:ABS}, the objective is to maximize the friction force applied by an antilock braking system (ABS) to a wheel with sensor noise, which extends the example considered in \cite[ch.~7]{KrsticBookESC2003} to include sensor noise.
%

\begin{exam}\label{ex:static_quadratic}
{\it Quadratic Cost.}
%
Consider 
\begin{align}
y(t) = \tfrac{1}{4} u^2(t),
\label{eq:cost_func}
\end{align}
where $u\in \BBR$ and $y \geq 0$.
The objective is to minimize $y$ by modulating $u$ in the presence of sensor noise $v,$ such that, for all $k \ge0,$
\begin{equation}
    v_k = \begin{cases}
   0.5  \sigma_k, & k \in [0, 1500], \\
   \sigma_k, & \mbox{otherwise},
    \end{cases} \label{ex1:v}
\end{equation}
where $\sigma_k\in\BBR$ is a Gaussian random variable with mean 0 and standard deviation $1;$ 
the sensor noise $v$ is shown in Figure \ref{fig:ex1_noise}.
For all  simulations in this example, the initial conditions are given by $u(0) = 10$ and $y(0) = 25.$
Since the system is a static map, the sampling rate is chosen to be $T_\rms = 1$ s without loss of generality and the results are presented in terms of steps.

\begin{figure}[h!]
    \centering
    \includegraphics[width=\columnwidth]{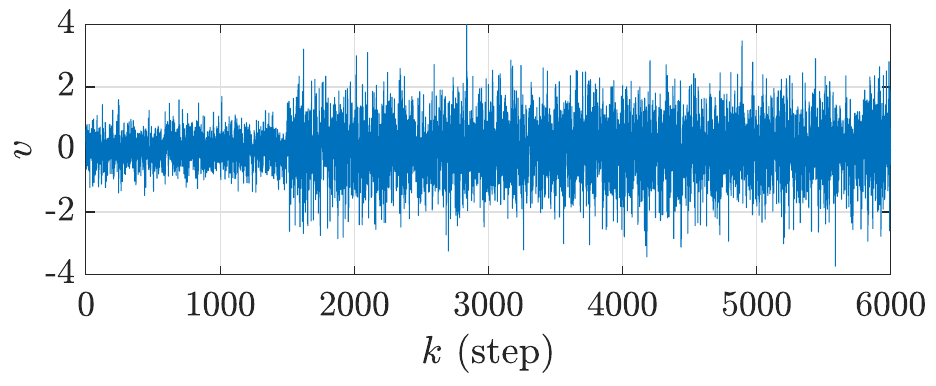}
    \caption{Example \ref{ex:static_quadratic}: {\bf Quadratic Cost}. Sensor noise $v$ defined in \eqref{ex1:v} added to $y$ for $k \in [0, 6000]$. Note that at $k = 1500$ step, the sensor noise level increases, thus introducing dynamic noise.} 
    \label{fig:ex1_noise}
\end{figure} 

The parameters for ESC are given by $K_\rmg = 1$, $K_{\rm esc} = -1.5$, $K_{{\rm en}, k} = 1$, $\omega_{\rm esc} = \pi/4$ rad/step/s, $A_{\rm esc} = 0.2$, $\omega_\rml = 2\pi/1000$ rad/step/s, $\omega_\rmh = 2\pi/100$ rad/step/s, and $u_0 = u(0) = 10$.
For ESC/AISE, the parameters are identical to those of ESC, with the exception of $\omega_\rmh,$ which is not used.
Additionally, the parameters for AISE are given by $n_\rme = 1$, $n_\rmf = 2$, $R_z = 1$, $R_d = 10^{-8}$, $R_\theta = 10^{-7}I_{3}$, $\eta = 0.02$, $\tau_n = 5$, $\tau_d = 25$, $\alpha = 0.02$, and $R_{\infty} = 10^{4},$
%
$\eta_{\rmL} = 10^{-6}$, $\eta_{\rmU} = 1$, and $\beta = 0.5.$
%
For RMSE calculation, we set $k_{\rm init} = 2000,$ $k_{\rm end} = 6000,$ and $u_{{\rm opt},k} = 0$ since this value minimizes $y.$

%
%
Figures \ref{fig:ex1_y} and \ref {fig:ex1_u} show the results of implementing ESC and ESC/AISE on the quadratic cost given by \eqref{eq:cost_func} with the sensor noise $v$ shown in Figure \ref{fig:ex1_noise}, which shows that, while both methods drive $u$ to a neighborhood of the minimizer 0, for all $k \ge 1500,$ the disruptions of $u$ due to sensor noise are less visible in the response of ESC/AISE, which shows that the performance of ESC/AISE is less degraded by sensor noise.

\begin{figure}[h!]
    \centering
    \includegraphics[width=\columnwidth]{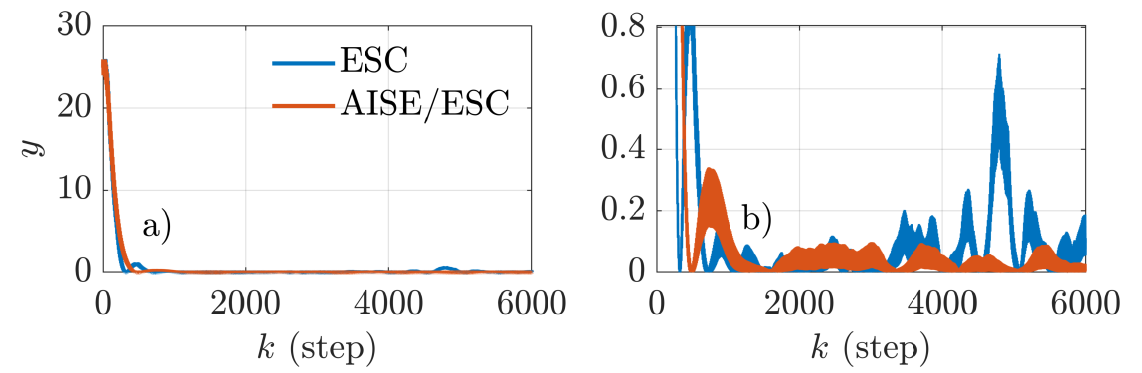}
    \caption{Example \ref{ex:static_quadratic}: {\bf Quadratic Cost}. System output $y$ for the quadratic cost given by \eqref{eq:cost_func} using ESC and ESC/AISE with the sensor noise $v$ shown in Figure \ref{fig:ex1_noise}.
    %
    %
    b) shows a) for all $y \in [0, 0.8].$} 
    \label{fig:ex1_y}
\end{figure} 

\begin{figure}[h!]
    \centering
    \includegraphics[width=\columnwidth]{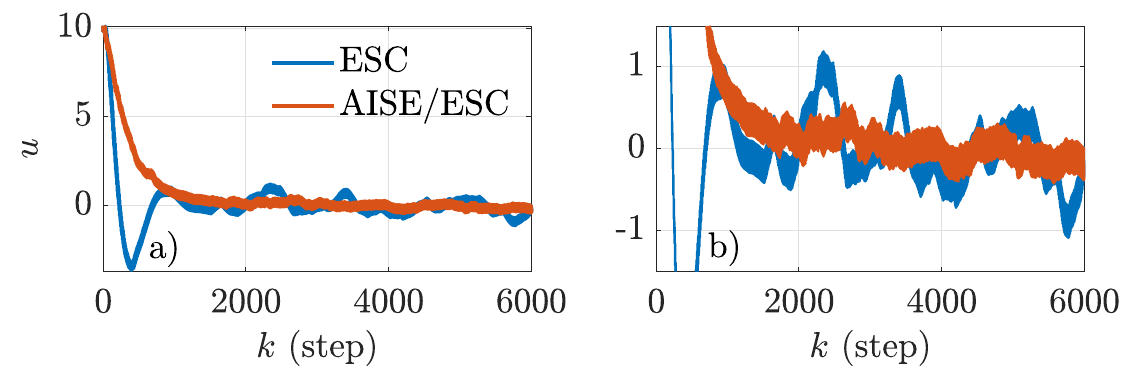}
    \caption{Example \ref{ex:static_quadratic}: {\bf Quadratic Cost}. Control input $u$ for the quadratic cost given by \eqref{eq:cost_func} using ESC and ESC/AISE with the sensor noise $v$ shown in Figure \ref{fig:ex1_noise}.
    b) shows a) for all $u \in [-1.5, 1.5].$} 
    \label{fig:ex1_u}
\end{figure}

To further test the sensitivity of ESC and ESC/AISE to sensor noise, 200 random trials are performed.
The average RMSE values computed over the 200 trials 
%
%
are shown in Table \ref{tab:ex1_MC_RMSE}, which suggests that, in the presence of sensor noise, ESC/AISE has overall better performance.
\hfill {\LARGE$\diamond$}

\begin{table}[h!]
\centering
\renewcommand{\arraystretch}{1.2}
\resizebox{0.6\columnwidth}{!}{%
\begin{tabular}{ |l||c|c| }
\hline
 \mbox{\bf Method}  & ESC & ESC/AISE  \\
 \hline
 \mbox{\bf Average RMSE} &  $0.476$ &   $0.240$ \\
 \hline
\end{tabular}
}
\vspace{0.5em}
\caption{Example \ref{ex:static_quadratic}: {\bf Quadratic Cost}. Average RMSE for ESC and ESC/AISE from 200 Monte Carlo trials.}
\label{tab:ex1_MC_RMSE}
%
\end{table}

\end{exam}

\begin{exam}\label{ex:ABS}
{\it Antilock Braking System (ABS).}
Consider an ABS implemented in a single-wheel system with dynamics 
%
\begin{align}
\dot{\nu} &= -g \mu_\lambda, \label{eq:ex_dotv}\\
\dot{\Omega} &= -\frac{B_\rmf}{J_\rmw} \Omega + \frac{m g R}{J_\rmw} \mu_\lambda - \tau_\rmB, \label{eq:ex_dotOmega}
\end{align}
where $\nu\in\BBR$ is the forward velocity of the center of the wheel, $\Omega\in\BBR$ is the angular velocity of the wheel, $m, R, J_\rmw$ are the mass, radius, and moment of inertia of the wheel, respectively, $B_\rmf$ is the bearing friction torque coefficient, $g$ is the acceleration due to gravity, $\tau_\rmB$ is the braking torque, $\lambda > 0$ is the wheel slip defined as
\begin{equation*}
    \lambda \isdef \frac{\nu - R \Omega}{v},
\end{equation*}
where $\dot{\nu}<0$ and $R\Omega \le \nu,$ and $\mu_\lambda$ is the friction force coefficient for all $\lambda > 0.$
For  simulation, $\mu_\lambda$ is defined as
\begin{equation}
\mu_\lambda \isdef 2 \mu^\star \frac{\lambda^\star \lambda}{(\lambda^\star)^2 + \lambda^2}, \label{eq:ex2_lambda_mu}
\end{equation}
where $\mu^\star$ is the maximum value of $\mu_\lambda$ and $\lambda^\star$ is the maximizer of $\mu_\lambda,$ such that $\mu_{\lambda^\star} = \mu^\star,$ as shown in Figure \ref{fig:ex2_lambda_mu}.

\begin{figure}[h!]
    \centering
    \includegraphics[width=\columnwidth]{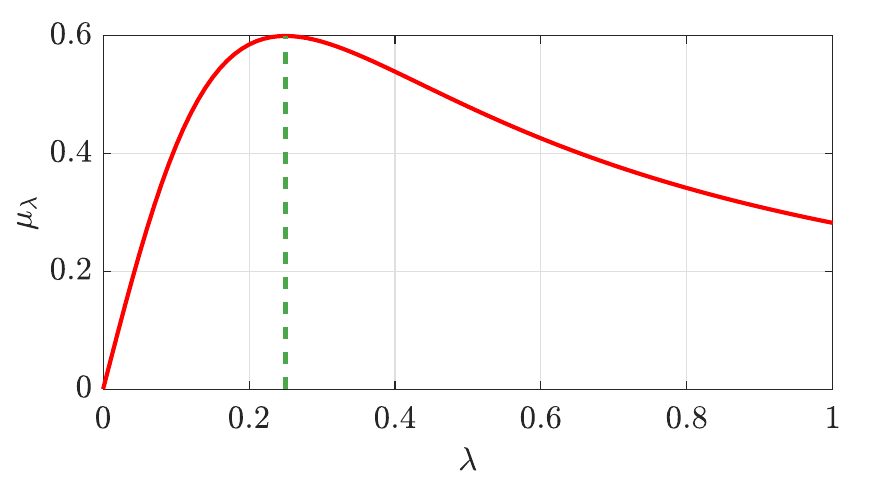}
    \caption{Example \ref{ex:ABS}: {\bf Antilock Braking System}. $\mu_\lambda$ versus $\lambda$ given by \eqref{eq:ex2_lambda_mu} for $\lambda^\star = 0.25$ and $\mu^\star = 0.6.$ The vertical, dashed green line indicates the value at which $\lambda = \lambda^\star,$ which crosses the $\mu_\lambda$ versus $\lambda$ trace at $\mu_{\lambda^\star} = \mu^\star,$ which is its maximum value.} 
    \label{fig:ex2_lambda_mu}
\end{figure} 

We assume that $\dot{\nu}$ is measured by an accelerometer and $\tau_\rmB$ given by the feedback linearizing controller
\begin{equation}
    \tau_\rmB = - \frac{c J_\rmw \nu}{R} (\lambda - \lambda_\rmd) - B_\rmf \Omega - \frac{J_\rmw \Omega}{\nu} \dot{\nu} - m R \dot{\nu},\label{eq:ex_taub}
\end{equation}
where $c > 0$ and $\lambda_\rmd > 0$ is the designated value of $\lambda.$ 
Note that, when $\lambda_\rmd$ is constant,  \eqref{eq:ex_taub} and \cite[p.~94, (7.4)]{KrsticBookESC2003} imply that, for all $t \ge 0,$ $\tilde{\lambda} (t) = \tilde{\lambda}(0) e^{-ct},$ where $\tilde{\lambda} \isdef \lambda - \lambda_\rmd.$

The closed-loop system consisting of the single-wheel system with the ABS and the feedback linearizing controller is given by \eqref{eq:ex_dotv}--\eqref{eq:ex_taub}.
The objective of the ABS is to maximize the stopping rate of the wheel, which is accomplished by reaching a value of $\lambda$ that maximizes $\mu_\lambda,$ as shown by \eqref{eq:ex_dotv}.
Since the controller given by \eqref{eq:ex_taub} modulates $\lambda_\rmd$ to reach a designated value of $\lambda,$ the objective is to determine $\lambda_\rmd$ such that $\mu_\lambda$ is maximized.
Hence, the objective is to maximize $y \isdef \mu_\lambda$ by modulating $u \isdef \lambda_\rmd$ in the presence of sensor noise $v$ such that, for all $k \ge0,$
\begin{equation}
    v_k = \begin{cases}
    0.375 \ \sigma_k, & k \in [0, 1250], \\
    0.75 \ \sigma_k, & \mbox{otherwise},
    \end{cases} \label{ex2:v}
\end{equation}
where $\sigma_k$ is a Gaussian random variable with mean 0 and standard deviation $1;$ 
the sensor noise $v$ is shown in Figure \ref{fig:ex2_noise}.

\begin{figure}[h!]
    \vspace{1em}
    \centering
    \includegraphics[width=\columnwidth]{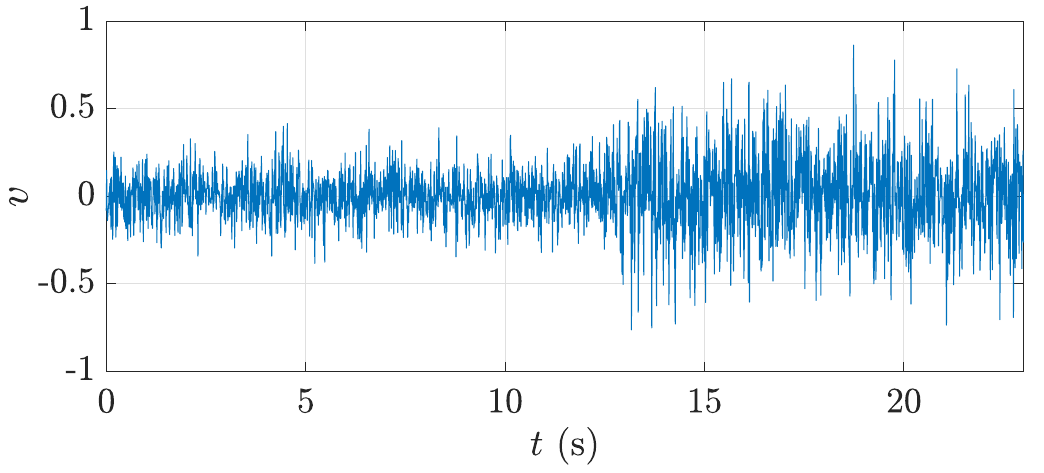}
    \caption{Example \ref{ex:ABS}: {\bf Antilock Braking System}. Sensor noise $v$ defined in \eqref{ex2:v} added to $y$ for $t \in [0, 23]$ s. Note that at $t = 13$ s, the sensor noise level increases, thus introducing dynamic noise.} 
    \label{fig:ex2_noise}
\end{figure}

Furthermore, for all  simulations in this example, the wheel and feedback linearizing controller parameters are given by $m = 400$ kg, $J_\rmw = 1$ kg $\cdot$ m$^2,$ $R = 0.3$ m,  $B_\rmf = 0.01$ kg $\cdot$ m$^2$/s, $\lambda^\star = 0.25,$ $\mu^\star = 0.6,$ and $c = 2,$ the initial conditions are given by $\lambda_\rmd(0) = 0.1,$ $\nu(0) = 336/3.6$ m/s and $\Omega(0) = 1120/3.6$ rad/s, such that $\lambda(0) = 0,$ and the sampling rate is given by $T_\rms = 0.01$ s.
The continuous-time dynamics are simulated in Matlab by using ode45 with simulation time step 0.01 s. %
The simulation finalizes when the wheel stops, that is, $\nu$ reaches 0, or a maximum time limit of 50 s is reached. %
The time-to-stop $t_{\rm stop} > 0$, which is defined by
%
\begin{equation*}
    t_{\rm stop} \isdef \min \{t > 0 \colon \nu(t) = 0 \},
\end{equation*}
%
is used as the performance variable since the underlying objective of this problem is to stop the wheel as quickly as possible.

The parameters for ESC are given by $K_\rmg = 1$, $K_{\rm esc} = 1500$, $K_{{\rm en}, k} = 1$, $\omega_{\rm esc} = 10$ rad/step/s, $A_{\rm esc} = 0.01$, $\omega_\rml = 8$ rad/step/s, $\omega_\rmh = 6$ rad/step/s, and $u_0 = \lambda_\rmd(0) = 0.1.$
For ESC/AISE, the parameters are identical to those of ESC, with the exception that the $\omega_\rmh$ parameter is no longer used.
Additionally, for AISE are given by $n_\rme = 10$, $n_\rmf = 20$, $R_z = 1$, $R_d = 10^{1}$, $R_\theta = 10^{-2}I_{3}$, $\eta = 0.001$, $\tau_n = 2$, $\tau_d = 10$, $\alpha = 0.02$, and $R_{\infty} = 10^{4},$
%
$\eta_{\rmL} = 10^{-8}$, $\eta_{\rmU} = 10^4$, and $\beta = 0.55.$
%
For RMSE calculation, we set $k_{\rm init} = 5 / T_\rms,$ $k_{\rm end} = \lfloor\min\{t_{\rm stop}, 50\} / T_\rms \rfloor,$ and $u_{\rm opt} = \lambda^\star$ due to the maximum time limit of 50 s. %
%

Figures \ref{fig:ex2_esc_noiseless} and \ref {fig:ex2_aise_esc_noiseless} and Table \ref{tab:ex2_RMSE_no_noise} show the results of implementing ESC and ESC/AISE on the ABS dynamics given by \eqref{eq:ex_dotv}--\eqref{eq:ex_taub} under no sensor noise, such that $v \equiv 0.$
These results show that ESC and ESC/AISE have similar maximization performance when no sensor noise is added to the ABS system measured output.

\begin{figure}[h!]
    \vspace{1em}
    \centering
    \includegraphics[width=\columnwidth]{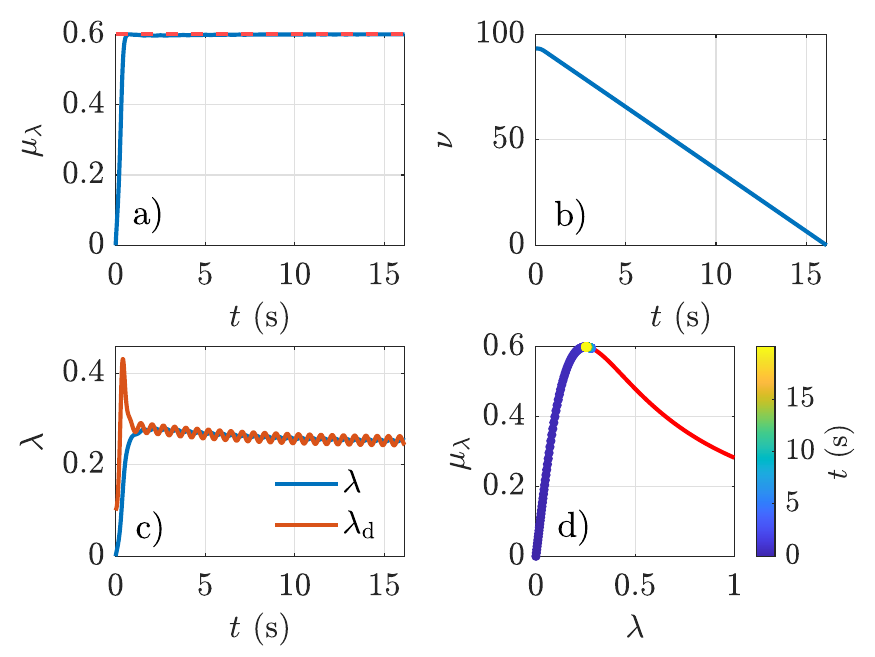}
    \caption{Example \ref{ex:ABS}: {\bf Antilock Braking System}. Results of implementing discrete-time ESC on the ABS dynamics given by \eqref{eq:ex_dotv}--\eqref{eq:ex_taub} under no sensor noise, such that $v \equiv 0.$
    a) shows $\mu_\lambda$ versus time, where the horizontal, dashed red line indicates the optimal value of $\mu$ given by $\mu^\star.$
    b) shows $\nu$ versus time.
    c) shows $\lambda$ versus time and $\lambda_\rmd$ versus time.
    d) shows $\mu_\lambda$ versus $\lambda$ over time, where the red curve corresponds to $\mu_\lambda$ versus $\lambda$ given by \eqref{eq:ex2_lambda_mu} and shown in Figure \ref{fig:ex2_lambda_mu}.
    } 
    \label{fig:ex2_esc_noiseless}
\end{figure}

\begin{figure}[h!]
    \centering
    \includegraphics[width=\columnwidth]{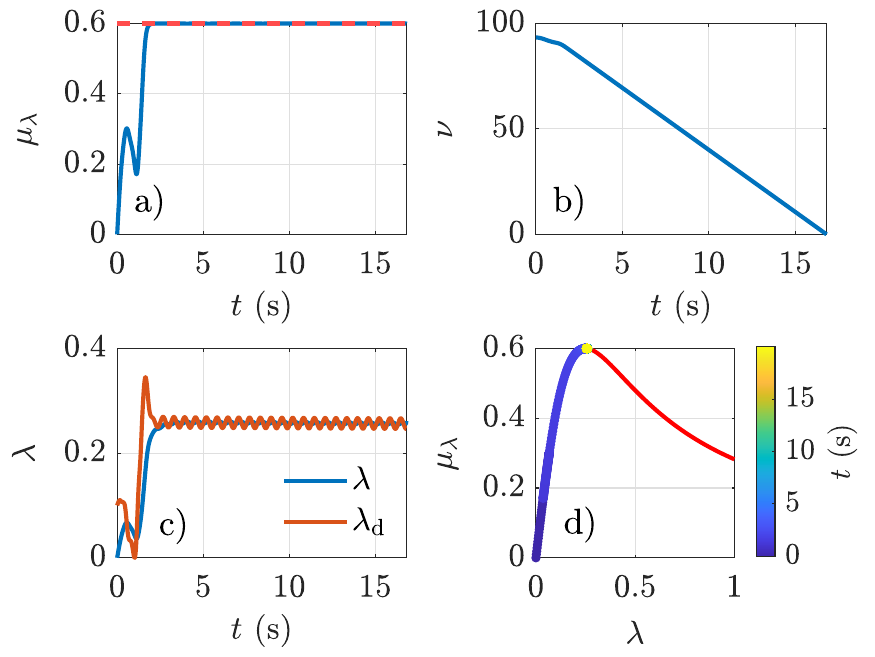}
    \caption{Example \ref{ex:ABS}: {\bf Antilock Braking System}. Results of implementing ESC/AISE on the ABS dynamics given by \eqref{eq:ex_dotv}--\eqref{eq:ex_taub} under no sensor noise, such that $v \equiv 0.$
    a) shows $\mu_\lambda$ versus time, where the horizontal, dashed red line indicates the optimal value of $\mu$ given by $\mu^\star.$
    b) shows $\nu$ versus time.
    c) shows $\lambda$ versus time and $\lambda_\rmd$ versus time.
    d) shows $\mu_\lambda$ versus $\lambda$ over time, where the red curve corresponds to $\mu_\lambda$ versus $\lambda$ given by \eqref{eq:ex2_lambda_mu} and shown in Figure \ref{fig:ex2_lambda_mu}.} 
    \label{fig:ex2_aise_esc_noiseless}
\end{figure}

\begin{table}[h!]
\vspace{1em}
\centering
\renewcommand{\arraystretch}{1.2}
\resizebox{0.55\columnwidth}{!}{%
\begin{tabular}{ |l||c|c| }
\hline
\textbf{Metric} & \textbf{ESC} & \textbf{ESC/AISE} \\ 
\hline
\textbf{RMSE} &  0.0080201 & 0.0074521 \\ 
\hline
$\bm{t_{\rm stop}}$ \mbox{\bf (s)} & 16.41 & 16.81 \\ 
\hline 
\end{tabular}
}
\vspace{0.5em}
\caption{Example \ref{ex:ABS}: {\bf Antilock Braking System}. RMSE and $t_{\rm stop}$ for application of ESC and ESC/AISE on the ABS system without sensor noise.}
\label{tab:ex2_RMSE_no_noise}
%
\end{table}

Next, Figures \ref{fig:ex2_esc_noisy} and \ref {fig:ex2_aise_esc_noisy} show the results of implementing ESC and ESC/AISE on the ABS dynamics given by \eqref{eq:ex_dotv}--\eqref{eq:ex_taub}  with the sensor noise $v$ shown in Figure \ref{fig:ex2_noise}.
These results show that, while both methods stop the wheel and drive $\nu$ to 0, the disruptions due to the added sensor noise are less visible in the response of ESC/AISE, which also yields a lower value of $t_{\rm stop},$ which shows that the performance of ESC/AISE is less degraded by sensor noise.

\begin{figure}[h!]
    \vspace{1em}
    \centering
    \includegraphics[width=\columnwidth]{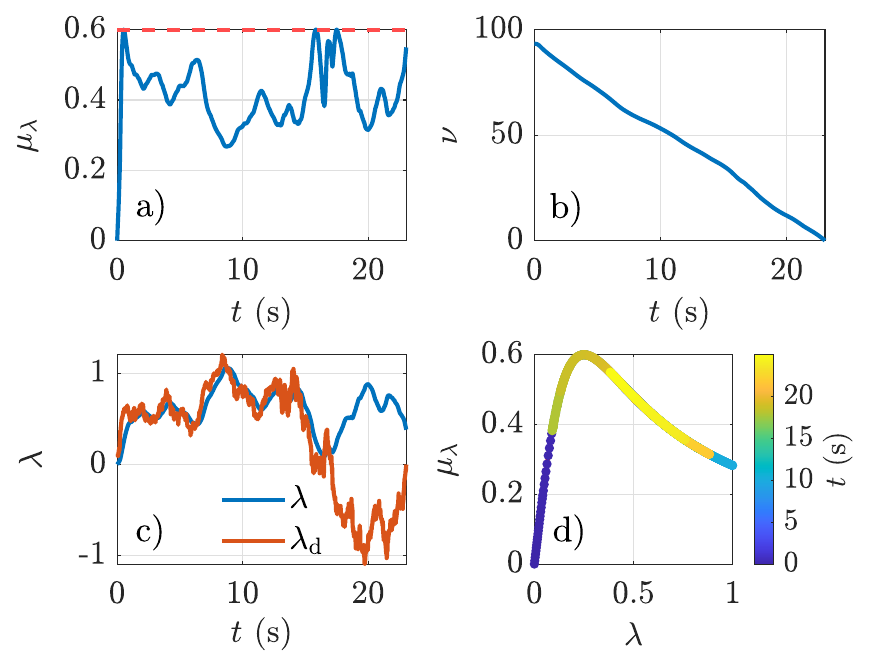}
    \caption{Example \ref{ex:ABS}: {\bf Antilock Braking System}. Results of implementing discrete-time ESC on the ABS dynamics given by \eqref{eq:ex_dotv}--\eqref{eq:ex_taub} with the sensor noise $v$ shown in Figure \ref{fig:ex2_noise}.
    a) shows $\mu_\lambda$ versus time, where the horizontal, dashed red line indicates the optimal value of $\mu$ given by $\mu^\star.$
    b) shows $\nu$ versus time.
    c) shows $\lambda$ versus time and $\lambda_\rmd$ versus time.
    d) shows $\mu_\lambda$ versus $\lambda$ over time, where the red curve corresponds to the $\mu_\lambda$ versus $\lambda$ given by \eqref{eq:ex2_lambda_mu} and shown in Figure \ref{fig:ex2_lambda_mu}.} 
    \label{fig:ex2_esc_noisy}
\end{figure} 

\begin{figure}[h!]
    \centering
    \includegraphics[width=\columnwidth]{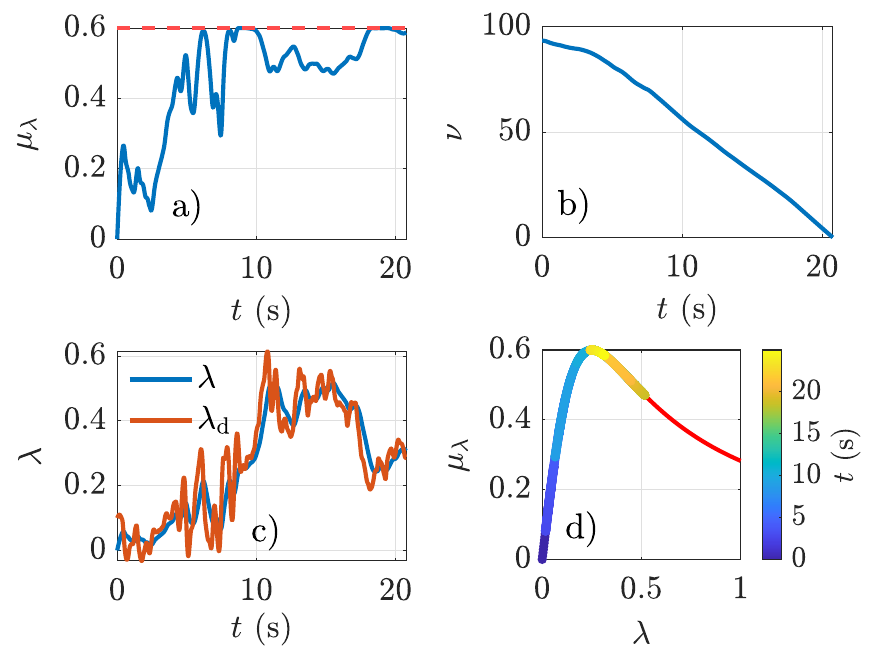}
    \caption{Example \ref{ex:ABS}: {\bf Antilock Braking System}. Results of implementing ESC/AISE on the ABS dynamics given by \eqref{eq:ex_dotv}--\eqref{eq:ex_taub} with the sensor noise $v$ shown in Figure \ref{fig:ex2_noise}.
    a) shows $\mu_\lambda$ versus time, where the horizontal, dashed red line indicates the optimal value of $\mu$ given by $\mu^\star.$
    b) shows $\nu$ versus time.
    c) shows $\lambda$ versus time and $\lambda_\rmd$ versus time.
    d) shows $\mu_\lambda$ versus $\lambda$ over time, where the red curve corresponds to $\mu_\lambda$ versus $\lambda$ given by \eqref{eq:ex2_lambda_mu} and Figure \ref{fig:ex2_lambda_mu}.} 
    \label{fig:ex2_aise_esc_noisy}
\end{figure}

To further test the sensitivity of ESC and ESC/AISE to sensor noise, 100 random trials are performed.
%
%
The average RMSE and average $t_{\rm stop}$ values computed over the 100 trials for the cases where ESC and ESC/AISE are implemented are shown in Table \ref{tab:ex1_MC_RMSE}.
A stopping performance metric is also included in Table \ref{tab:ex1_MC_RMSE}, which shows the percentage of runs in which $\nu$ is driven to 0 before the 50-s time limit.
These results demonstrate that ESC/AISE exhibits superior maximization performance and is more consistent in stopping the wheel within the time limit, even in the presence of sensor noise.
%
%
%
%
\hfill {\LARGE$\diamond$}
\begin{table}[h!]
\vspace{1em}
\centering
\renewcommand{\arraystretch}{1.2}
\resizebox{\columnwidth}{!}{%
\begin{tabular}{ |l||c|c| }
\hline
\textbf{Metric} & \textbf{ESC} & \textbf{ESC/AISE} \\ 
\hline
\mbox{\bf Average RMSE} & 2.0281 & 0.22208 \\ 
\hline
\mbox{\bf Average} $\bm{t_{\rm stop}}$ \mbox{\bf (s)} & 30.3806 & 24.3548 \\ 
\hline
\hspace{-0.6em}$\begin{array}{l} \mbox{\bf Stopping} \\ \mbox{\bf performance} \end{array}$ & \makecell{Stopped 64\% of the \\trials within 50 s} & \makecell{Stopped 100\% of the \\trials within 50 s} \\
\hline 
\end{tabular}
}
\vspace{0.5em}
\caption{Example \ref{ex:ABS}: {\bf Antilock Braking System}. Average RMSE, average $t_{\rm stop},$ and stopping performance for application of ESC and ESC/AISE on an ABS system with sensor noise from 100 random trials.}
\label{tab:ex2_MC_RMSE}
\end{table}
\end{exam}

\section{Conclusions}\label{sec:conclusions}
This paper introduced extremum-seeking control with adaptive input and state estimation (ESC/AISE) to improve optimization in the presence of measurement noise.
ESC/AISE is constructed by replacing the high-pass filter in the discrete-time ESC with AISE, which performs numerical differentiation that is robust to sensor noise.
%
%
Numerical studies benchmarked against classical ESC show consistently lower error and faster convergence in noisy settings, as shown in Tables \ref{tab:ex1_MC_RMSE} and \ref{tab:ex2_MC_RMSE}.
Future work will extend ESC/AISE to MIMO systems and variations of ESC, such as Newton-based ESC \cite{ghaffari2012}.
%
%
%

%

\section*{Acknowledgments}
This research was supported in part by NSF grant CMMI 2031333.


\bibliographystyle{IEEEtran}
\bibliography{bib_paper,bibpaper_sha}

\end{document}